\documentclass[10pt,twocolumn,nofootinbib]{revtex4}
\usepackage{amsmath,subfigure, graphicx, tabls, psfrag, dcolumn}

\begin{document}

\title{Transverse Emittance Dilution due to Coupler Kicks in Linear Accelerators}
\author{Brandon Buckley, Georg H. Hoffstaetter\footnote{Electronic Address: Georg.Hoffstaetter@cornell.edu}}
\affiliation{Laboratory for Elementary Particle Physics, Cornell University, Ithaca, NY, 14853, USA}

\begin{abstract}
One of the main concerns in the design of low emittance linear accelerators 
(linacs) is the 
preservation of beam emittance. Here we discuss one possible source of 
emittance dilution, the coupler kick, due to transverse electromagnetic fields in the 
accelerating cavities of the linac caused by the power coupler geometry.
In addition to emittance growth, the coupler kick also produces orbit distortions.
It is common wisdom that emittance growth from coupler kicks can 
be strongly reduced by using two couplers per cavity mounted opposite each other 
or by having the couplers of successive cavities alternation from above to below the beam pipe
so as to cancel each individual kick.
While this is correct, including two couplers per cavity or alternating the coupler location
requires large technical changes and increased cost for superconducting cryomodules where cryogenic pipes are
arranged parallel to a string of several cavities. We therefore analyze consequences of
alternate coupler placements.

We show here that alternating the coupler location from above to below
compensates the emittance growth as well as the orbit distortions. And
for sufficiently large Q values, alternating the coupler location from
before to after the cavity leads to a cancellation of the orbit
distortion but not of the emittance growth, whereas alternating the
coupler location from before and above to behind and below the cavity
cancels the emittance growth but not the orbit distortion. We show
that cancellations hold for sufficiently large Q values.  These
compensations hold even when each cavity is individually detuned,
e.g. by microphonics.  Another effective method for reducing coupler
kicks that is studied is the optimization of the phase of the coupler
kick so as to minimize the effects on emittance from each coupler.
This technique is independent of the coupler geometry but relies on
operating on crest.  A final technique studied is symmetrization of
the cavity geometry in the coupler region with the addition of a stub
opposite the coupler.  This technique works by reducing the amplitude
of the off axis fields and is thus effective for off crest
acceleration as well.

We show applications of these techniques to the energy recovery linac
(ERL) planned at Cornell University.

\end{abstract}
\maketitle{}
\section{Introduction}

A possible source of emittance dilution in an accelerating cavity is that caused by the input power coupler.
The addition of a  
single coupler situated perpendicular to the beam pipe creates an 
asymmetry in the cavity geometry, leading to non radially symmetric field profiles
in the beam pipe in the vicinity of the coupler \cite{Zhang}.  The asymmetric fields produce a transverse radio 
frequency (rf) kick to an accelerating bunch resulting in an increase in emittance \cite{Dohlus}.  Additionally, 
the coupler region will change the 
cavities' RF focusing somewhat because the transverse dependence of 
fields in this region is different to that in the main cavity. This 
effect changes the emittance growth due to cavity focusing, but it is 
not considered part of the coupler kick and is not discussed here.  Previous studies 
have found the effect on emittance due to the transverse rf kick 
to be significant in the injector cavities of the Cornell energy recovery linac 
(ERL) \cite{Belomestnykh02,shemelin,Greenwald}.  As a solution, a second input coupler was installed situated
on the opposite side of the beam pipe, canceling the asymmetry and the transverse kick \cite{shemelin3}.  This approach,
though effective, would be both a technically challenging and expensive design for a large superconducting linac such 
as the Cornell ERL or the ILC.  
A solution to the emittance increase 
due to coupler kicks that does not include the addition of a second coupler would therefore be preferable.

In this paper we investigate the effects from a transverse rf-coupler kick on the emittance of a Gaussian bunch and
discuss possible methods of reducing emittance growth.  We consider and compare the effects from six different coupler configurations: 
(tf) all couplers mounted on the top of the beam pipe; all couplers placed in front of the cavity, 
(ta) all couplers mounted on the top of the beam pipe;  couplers alternated from being placed in front of and behind the cavity each cavity, 
(af) couplers alternated from being mounted on top of and underneath the beam pipe each cavity; all couplers placed in front of the cavity,
(aa) couplers alternated from being mounted on top of and underneath the beam pipe each cavity; 
couplers alternated from being placed in front of and behind the cavity each cavity,
(mf) couplers alternated from being mounted on top of and underneath the beam pipe each cryomodule, or every ten cavities; 
all couplers placed in front of the cavity,
(dc) double coupler arrangement with two couplers per cavity, equivalent to no transverse kick. 
The proposed design for the Cornell ERL includes alternating the coupler placement from in front of and 
behind the cavity each cavity, 
as in configurations (ta) and (aa).  The configurations (tf) and (af) are included for comparison so 
as to investigate the effects 
from alternating the placement of the coupler from front to back.  The (mf) configuration is included 
so as to investigate the extent of the
cancellation between two cryomodules.  
Of the two configurations (ta) and (aa) the most preferable would be configuration (ta) as it includes mounting 
couplers all on the same side of the beam
pipe and is thus technically more feasible. 
In addition to these six configurations we investigate the effects due to optimizing the placement of 
the coupler along the beam pipe
and the effects due to the addition of a symmetrizing stub opposite the coupler.

\begin{table}[tp]
\caption{Parameters of accelerating cavities for the Cornell ERL.
\label{cavparams}}
\begin{ruledtabular}
\begin{tabular}{lccc}
Frequency & 1300 MHz \\
Number of Cells& 7 \\ 
Cavity Shape & TESLA type\\
Accelerating Voltage& 15 MV/m \\
$Q_0$&  $10^{10}$ \\
$Q_{ext}$ & $10^8$ \\
Coupler Type& Coaxial \\
Coax Impedance&50 $\Omega$ \\
\end{tabular}
\end{ruledtabular}
\end{table}

\begin{figure}[tbhp]
\includegraphics[width=\columnwidth, clip]{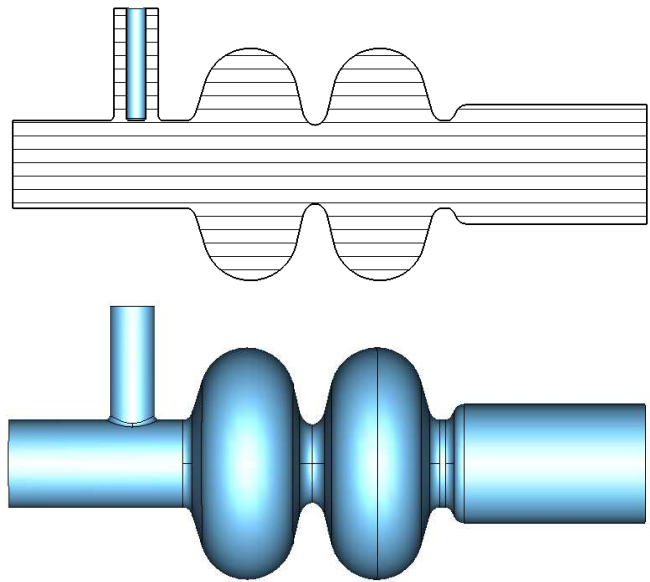}
\caption{Two cell model of the seven cell TESLA type Cornell ERL superconducting rf cavity.    
\label{figcav}}
\end{figure}

We simulate, using Microwave Studios (MWS) \cite{MWS}, the electric and magnetic standing wave profiles 
inside an accelerating cavity with
the coaxial coupler included (Fig. \ref{figcav}).  The cavity used for simulation is a two cell model of 
the seven cell TESLA-type cavity to 
be used in the proposed Cornell ERL.  A two cell cavity instead of a seven cell cavity is used in order 
to limit the simulation time.  From the standing
wave profiles of MWS, complex traveling waves are modeled of which the real parts represent the true waves in the cavity. 
A numerical integration of these waves is
performed along the central cavity axis to calculate the total change in momentum of a charged 
particle traveling through the cavity.  The coupler
kick, defined as the ratio of the transverse change in momentum and the change in momentum along 
the cavity axis, is calculated and input 
into a lattice representing the proposed Cornell ERL.  A simulation of an electron bunch through 
the lattice is done with 
BMAD \cite{sagan} and the total normalized
emittance growth is calculated and compared for all mentioned configurations.

We find that due to the high $Q_{ext}$ values of the accelerating cavities, the fields on the cavity axis, including 
those in the vicinity of the coupler, are very well approximated by standing waves.  From this 
approximation we formulate analytical 
arguments to support the results from our simulation, namely that the orbit distortion is canceled.
Furthermore,
from the standing wave approximation, we present arguments to back up the results from simulations indicating that 
the coupler kick is independent of reflected waves in the coupler and of relative phase differences between incoming
and reflected waves.  Thus our result of the cancellation of the coupler kick between adjacent cavities
is unaffected by cavity detuning.  

Lastly, we 
show that placing the coupler at a distance from the entrance of the cavity so as
to match the phases of the coupler kick and accelerating kick minimizes the emittance increase,
as does the addition of a symmetrizing stub which effectively minimizes the amplitudes of the off axis fields
in the beam pipe.
This additionally minimizes the orbit distortion.
Important to note is that emittance growth due to higher order mode (HOM) couplers
can be dealt with using all of the above techniques in an analogous way.
  
The Linac parameters used for simulations of the Cornell ERL are listed in Table \ref{cavparams}.

\section{Emittance Growth due to Coupler Kick}

In this section an analytical expression is derived for the change in emittance of a relativistic, Gaussian distributed bunch due 
to a transverse rf kick in an accelerating
rf cavity.  We begin by defining the change in transverse momentum, in this case the y component:
\begin{equation}
\Delta P_y=\frac{\Delta E_0}{c} |\kappa| e^{i[\phi_c+\psi+\omega(t-t_0)]} \ , \ \ \Delta p_y=Re\{\Delta P_y\}.
\label{eq1}
\end{equation}
In the above, $Re\{(\Delta E_0/c)e^{i[\psi+\omega(t-t_0)]}\}$ is the change in momentum in the longitudinal direction, $p_s=Re\{\Delta P_s\}$, 
for a particle at an offset $\Delta t= t-t_0$ from 
the center of the bunch.  
The coupler kick $\kappa$ is defined as the ratio of the complex transverse rf kick with the complex longitudinal kick \cite{Dohlus2}:
\begin{equation}
\kappa=\frac{\Delta P_y}{\Delta P_s}
\label{eqkappa}
\end{equation}
The phase of the coupler kick, $\phi_c$, is  
the difference between the phase of the the transverse kick and $\psi$, the phase
of the accelerating kick with respect to the reference particle at the center of the bunch.  
Dividing by the initial longitudinal momentum $E/c$ we achieve the change in the phase space component $y'$:
\begin{equation}
\Delta y'=Re\{\frac{\Delta E_0}{E} |\kappa| e^{i(\phi_c+\psi+\omega\Delta t)} \}.
\label{eq2}
\end{equation}
Expanding to first order in $\Delta t$ leads to the approximate expression
\begin{align}
\label{eq3}
\Delta y'&\approx Re\{\frac{\Delta E_0}{E} |\kappa|e^{i(\phi_c+\psi)}(1+i\omega\Delta t)\} \\
         &\approx\frac{\Delta E_0}{E}|\kappa|\{\cos(\phi_c+\psi)-\omega\sin(\phi_c+\psi)\Delta t\}\nonumber \\
         &\approx\Delta y_0'-S\Delta t, \nonumber
\end{align}
with $\Delta y_0'=\frac{\Delta E_0}{E} |\kappa|\cos(\phi_c+\psi)$ and $S=\frac{\Delta E_0}{E} |\kappa|\omega\sin(\phi_c+\psi)$.

From $\Delta y'$ we are now able to deduce the change in emittance.  Beginning with a Gaussian distribution of particles defined by
\begin{equation}
\rho_0(y,y',\Delta t)=\frac{1}{2\pi\varepsilon_{y,0}}e^{-\frac{\gamma y^2+2\alpha yy'+\beta y'^2}{2\varepsilon_{y,0}}}\frac{1}{\sqrt{2\pi}\sigma_t}e^{-\frac{\Delta t^2}{2\sigma_t^2}}
\label{eq4}
\end{equation}
we can introduce the change in $y'$ of Eq.~(\ref{eq3}) ignoring, however, the constant change $\Delta y_0'$ term which must
be compensated for with orbit correctors.  The expression for $\rho$ in Eq.~(\ref{eq4}) then becomes
\begin{align}
\label{eqa}
\rho(y,y',\Delta t)=&\frac{1}{2\pi\varepsilon_{y,0}}e^{-\frac{\gamma y^2+2\alpha y(y'-S\Delta t)+\beta(y'-S\Delta t)^2}{2\varepsilon_{y,0}}}\\
&\times \frac{1}{\sqrt{2\pi}\sigma_t}e^{-\frac{\Delta t^2}{2\sigma_t^2}}. \nonumber
\end{align}
The final emittance
is given by
\begin{align}
\label{eq6}
\varepsilon_y&=\int(\frac{1}{2}(\gamma y^2 +2\alpha yy'+\beta y'^2)\rho(y,y',\Delta t)dydy'd\Delta t \\
             &=\varepsilon_{y,0}+\frac{1}{2}\beta S^2\sigma_t^2.\nonumber
\end{align}

\section{Synthesis of Standing Wave Patterns into Traveling Waves}

We use MWS to simulate standing electromagnetic field patterns inside
the accelerating cavity which can be chosen to satisfy a set of
boundary conditions at the end of the coupler: perfect electric wall,
for which there is no component of the electric field parallel to the
boundary, and perfect magnetic wall, for which there is no magnetic
field component parallel to the boundary.  We will henceforth refer to
this boundary surface as the coupler boundary.  The energy in the
resulting field patterns, $\mathbf{E}^e(\mathbf{r}),
\mathbf{B}^e(\mathbf{r}),\mathbf{E}^m(\mathbf{r}),\mathbf{B}^m(\mathbf{r})$,
for which the superscripts indicate the boundary condition, are
normalized to one Joule by MWS.  We choose the overall signs of the
fields such that $\mathbf{E}^m(z) \cdot \mathbf{e}_r, \mathbf{B}^e(z)
\cdot \mathbf{e}_{\phi},\frac{\partial}{\partial z}\mathbf{E}^e(z)
\cdot \mathbf{e_r}$ and $\frac{\partial}{\partial z}\mathbf{B}^m(z)
\cdot \mathbf{e_{\phi}}$ are all positive at the boundary of the
coupler, thus representing positive sines and cosines.  The
cylindrical coordinate system here is set up with the z axis pointing
down the axis of the coupler towards the entrance into the cavity.
Multiplying $\mathbf{E}^m(\mathbf{r})$ and $\mathbf{B}^m(\mathbf{r})$
by $\xi=c\mathbf{B}^e(0)\cdot
\mathbf{e}_{\varphi}/\mathbf{E}^m(0)\cdot \mathbf{e}_r$ will normalize
the amplitudes of these magnetic boundary condition fields inside the
coupler to the amplitudes of the corresponding electric boundary
condition fields.

Inside the coaxial coupler the standing 
wave patterns are then given by:  
\begin{align}
\label{eq8}
&\mathbf{E}^e(\mathbf{r})=\mathbf{e}_r \frac{A}{r}\sin(kz), \mathbf{B}^e(\mathbf{r})=\mathbf{e}_{\varphi} \frac{1}{c}\frac{A}{r}\cos(kz),\\ 
&\mathbf{E}^m(\mathbf{r})=\mathbf{e}_r \frac{A}{\xi r}\cos(kz), \mathbf{B}^m(\mathbf{r})=\mathbf{e}_{\varphi} \frac{1}{c}\frac{A}{\xi r}\sin(kz). \nonumber
\end{align}

If we combine these fields via the following, we will obtain expressions for waves traveling down and up the coupler,
indicated by + and - respectively:
\begin{align}
\label{eq9}
&\mathbf{E}^{\pm}(\mathbf{r},t)=Re\{(\xi\mathbf{E}^m(\mathbf{r})\pm i\mathbf{E}^e(\mathbf{r}))e^{-i(\omega t-\phi^{\pm})}\},\\
&\mathbf{B}^{\pm}(\mathbf{r},t)=\pm Re\{(\mathbf{B}^e(\mathbf{r})\pm i\xi\mathbf{B}^m(\mathbf{r}))e^{-i(\omega t-\phi^{\pm})}\},\nonumber
\end{align}
where $\phi^{\pm}$ are arbitrary phases which we will later choose conveniently. 

\subsection{Standing Wave Approximation}

We now consider the case of the fields inside the cavity on the central axis denoted by a subscript 0:
$\mathbf{E}_0^e(s),\mathbf{B}_0^e(s),\mathbf{E}_0^m(s)$ and $\mathbf{B}_0^m(s)$, with the s axis pointing down the cavity.  
We will use the approximation that traveling waves
in the coax excite standing waves in the cavity.  Exact standing waves would be excited in the cavity if the energy leaving the cavity through the coupler per oscillation,
$\delta E$,
were zero.
Correspondingly, this standing wave approximation is very good if the energy loss per oscillation is much less than the the total energy $W$ stored in the cavity.
The ratio between these two energies is characterized by
\begin{equation}
Q_{ext}=\frac{2\pi W}{\delta E}=\frac{\omega W}{P}
\label{eqQ}
\end{equation}
where $\omega$ is the resonant frequency of the cavity and P is the power dissipated from the cavity through the coupler. 

Hence,
\begin{align}
\label{eq10a}
&\mathbf{E}_0^{\pm}(s,t)=Re\{(\xi\mathbf{E}_0^m(s)\pm i\mathbf{E}_0^e(s))e^{-i(\omega t-\phi^{\pm})}\},\\
&\mathbf{B}_0^{\pm}(s,t)=\pm Re\{(\mathbf{B}_0^e(s)\pm i\xi\mathbf{B}_0^m(s))e^{-i(\omega t-\phi^{\pm})}\},\nonumber
\end{align}
should, to a good approximation, represent standing waves if $Q_{ext}$ is large.  As such, the fields should 
be products of a function of time and a function of s.   
The field pattern $\mathbf{E}_0^m(s)$ thus
must be proportional to $\mathbf{E}_0^e(s)$, as well as $\mathbf{B}_0^m(s)$ to $\mathbf{B}_0^e(s)$. Since 
the standing wave profiles are normalized to the same energy and since the energy inside the coupler can be deemed negligible
compared to the energy in the cavity, the proportionality constants must be of magnitude one and 
the fields on the s axis must be approximately equal up to a sign:
\begin{equation}
\mathbf{E}_0^e(s)\approx s^e\mathbf{E}_0^m(s), \mathbf{B}_0^e(s)\approx s^m\mathbf{B}_0^m(s)
\label{eq11}
\end{equation}
with $s^e,s^m\in\{-1,1\}$.

Substitution into Eq.~(\ref{eq10a}) leads to 
\begin{align}
\label{eq12}
&\mathbf{E}_0^{\pm}(s,t)\approx Re\{\mathbf{E}_0^m(s)(\xi\pm is^e)e^{-i(\omega t-\phi^{\pm})}\},\\
&\mathbf{B}_0^{\pm}(s,t)\approx \pm Re\{\mathbf{B}_0^m(s)(\pm i)(\xi \mp is^m)e^{-i(\omega t-\phi^{\pm})}\}.\nonumber
\end{align}
Now we choose $\phi^{\pm}$ such that $(\xi\pm i s^e)e^{i\phi^{\pm}}\in\Re$.  In order to satisfy Maxwell's equations we must then
also have $(\xi\mp i s^m)e^{i\phi^{\pm}}\in\Re$.  We therefore deduce that $s^m$ must equal $-s^e$ with $\phi^{\pm}=\pm s^e\cot^{-1}(\xi)$.
The waves in the cavity can thus be written as:
\begin{align}
\label{eq13}
&\mathbf{E}_0^{\pm}(s,t)\approx\mathbf{E}_0^m(s)A\cos(\omega t)\ ,\\
&\mathbf{B}_0^{\pm}(s,t)\approx\mathbf{B}_0^m(s)A\sin(\omega t)\ ,\nonumber
\end{align}
with
\begin{equation}
A=(\xi\pm i s^e)e^{i\phi^{\pm}}=\sqrt{\xi^2+1}
\label{eq13a}
\end{equation}
and
\begin{equation}
\mathbf{E}_0^e(s)\approx s^e\mathbf{E}_0^m(s), \mathbf{B}_0^e(s)\approx -s^e\mathbf{B}_0^m(s).
\label{eq14}
\end{equation}

\section{$Q_{ext}$ Considerations}
Even when the standing wave approximation is very good there is some region in the beam pipe, in the vicinity of the coupler, in which the traveling wave in the 
coax changes to a standing wave in the cavity.  This transition region will be smaller for larger $Q_{ext}$ and as such,
for very high $Q_{ext}$ values the waves excited on the cavity axis will be standing waves, even in the coupler region.  It is thus important to simulate in MWS a cavity
with the correct $Q_{ext}$ value in order to determine the accuracy of the standing wave approximation.  Factors in the geometry of a coaxial coupler affecting 
$Q_{ext}$ include the shape of the coupler, the distance from the entrance of the cavity and length of the inner conductor, i.e. the distance it penetrates into the 
beam pipe.    

\subsection{Calculating $Q_{ext}$} 

Several methods for calculating the external quality factor using computer codes have been described (\cite{Hartung,Balleyquier,Balleyquier2,shemelin2,Kroll}).
Below we derive an alternative method for calculating $Q_{ext}$ that utilizes the synthesized waves introduced in Section III. 

We begin by computing the total stored energy in the cavity  
via integration of the squares of the electric or magnetic fields over the entire cavity volume:
\begin{equation}
W=\frac{\varepsilon_0}{2} \int \!\!\!\! \int \!\!\!\! \int|\mathbf{\hat{E}}(\mathbf{r})|^2dv=\frac{1}{2\mu_0}\int \!\!\!\! \int \!\!\!\! \int|\mathbf{\hat{B}}
(\mathbf{r})|^2dv.
\label{eq39}
\end{equation}
In the above equation $\mathbf{\hat{E}}(\mathbf{r}) $ and $\mathbf{\hat{B}}(\mathbf{r})$ 
are complex field profiles of the oscillating electric and magnetic waves for which the real 
part is physical, i.e. $\mathbf{E}(\mathbf{r},t)=Re\{\mathbf{\hat{E}}(\mathbf{r})e^{-i\omega t}\}$ and 
$\mathbf{B}(\mathbf{r},t)=Re\{\mathbf{\hat{B}}(\mathbf{r})e^{-i\omega t}\}$. 
The power P dissipated through the coupler is found by taking the time average of the Poynting vector integrated over the coupler boundary: 
\begin{equation}
P=\frac{\varepsilon_0c}{2}\int \!\!\!\! \int |\mathbf{\hat{E}}(r,\varphi,0)|^2da=\frac{c}{2\mu_0}\int \!\!\!\! \int|\mathbf{\hat{B}}(r,\varphi,0)|^2da
\label{eq38}
\end{equation}
where $z=0$ signifies the coupler boundary. 
We now have two different expressions for $Q_{ext}$:
\begin{equation}
Q_{ext}=\frac{\omega\int \!\!\! \int \!\!\! \int|\mathbf{\hat{E}}(\mathbf{r})|^2dv}{c\int \!\!\! \int |\mathbf{\hat{E}}(r,\varphi,0)|^2da}=
\frac{\omega\int \!\!\! \int \!\!\! \int|\mathbf{\hat{B}}(\mathbf{r})|^2dv}{c\int \!\!\! \int |\mathbf{\hat{B}}(r,\varphi,0)|^2da}.
\label{eq40}
\end{equation}

We can now use our synthesized waves traveling up the coupler, $\mathbf{E}^-$ and $\mathbf{B}^-$ of Eq.~(\ref{eq9}) and insert them into our expression for $Q_{ext}$,
noting that in terms of the field profiles from MWS $\mathbf{\hat{E}}(\mathbf{r})=\xi\mathbf{E}^m(\mathbf{r})- i\mathbf{E}^e(\mathbf{r})$ and 
$\mathbf{\hat{B}}(\mathbf{r})=-(\mathbf{B}^e(\mathbf{r})- i\xi\mathbf{B}^m(\mathbf{r}))$:
\begin{align}
\label{eq41}
Q_{ext}&=\frac{\omega\int \!\!\! \int \!\!\! \int [\xi^2\mathbf{E}^m(\mathbf{r})^2+\mathbf{E}^e(\mathbf{r})^2]dv}{c\xi^2\int \!\!\! \int \mathbf{E}^m(r,\varphi,0)^2da}\\
       &=\frac{\omega\int \!\!\! \int \!\!\! \int [\xi^2\mathbf{B}^e(\mathbf{r})^2 +\mathbf{B}^m(\mathbf{r})^2]dv}{c\int \!\!\! \int \mathbf{B}^e(r,\varphi,0)^2da}.\nonumber
\end{align}
Due to the normalization of the energy in the cavity to one Joule in MWS the volume integrals are known: 
$\frac{\varepsilon_0}{2}\int \!\!\! \int \!\!\! \int \mathbf{E}^m(\mathbf{r})^2dv=\frac{\varepsilon_0}{2}
\int \!\!\! \int \!\!\! \int \mathbf{E}^e(\mathbf{r})^2dv=$ 1 J and 
$\frac{1}{2\mu_0}\int \!\!\! \int \!\!\! \int \mathbf{B}^m(\mathbf{r})^2dv=\frac{1}{2\mu_0}
\int \!\!\! \int \!\!\! \int \mathbf{B}^e(\mathbf{r})^2dv=$ 1 J.
  The surface integrals over the coupler boundary can be calculated with the 
knowledge of the field patterns in the coax from Eq.~(\ref{eq8}).  Inserting $z=0$ leaves the surface integral
\begin{align}
\label{eq42}
\int \!\!\!\! \int \mathbf{B}^e(r,\varphi,0)^2da&=\frac{\xi^2}{c^2}\int \!\!\!\! \int \mathbf{E}^m(r,\varphi,0)^2da \\
                                              &=\frac{A^2}{c^2}\int_0^{2\pi} \!\!\!\! \int_{r_i}^{r_o} \frac{1}{r} dr d\varphi\nonumber \\
                                              &=\frac{A^2}{c^2} 2\pi \ln\left(\!\frac{r_o}{r_i}\!\right).\nonumber
\end{align}
The amplitude $A$ can be found by taking a value of either the magnetic or electric 
field at an arbitrary radius, $r=a$, on the boundary, i.e. $A=\xi a|\mathbf{E}^m(a,\varphi,0)|=ca|\mathbf{B}^e(a,\varphi,0)|$.  Thus we have two equivalent expressions 
for $Q_{ext}$ requiring only two simulated values from MWS:
\begin{equation}
Q_{ext}=\frac{\xi^2+1}{\xi^2}\frac{\omega}{c\varepsilon_0\pi}\frac{1 {\rm J}}{a^2|\mathbf{E}^m(a,\varphi,0)|^2\ln(\frac{r_o}{r_i})}
\label{eq43}
\end{equation}
and
\begin{equation}
Q_{ext}=(\xi^2+1)\frac{\omega}{c\varepsilon_0\pi}\frac{1 {\rm J}}{a^2|c\mathbf{B}^e(a,\varphi,0)|^2\ln(\frac{r_o}{r_i})}.
\label{eq44}
\end{equation}
Since the cavity in our simulations is a 2 cell model of the actual 7 cell
ERL cavity, we multiplied these $Q_{ext}$ values by 3.5.  

\begin{figure*}[!thp]
\includegraphics[width=\textwidth, clip]{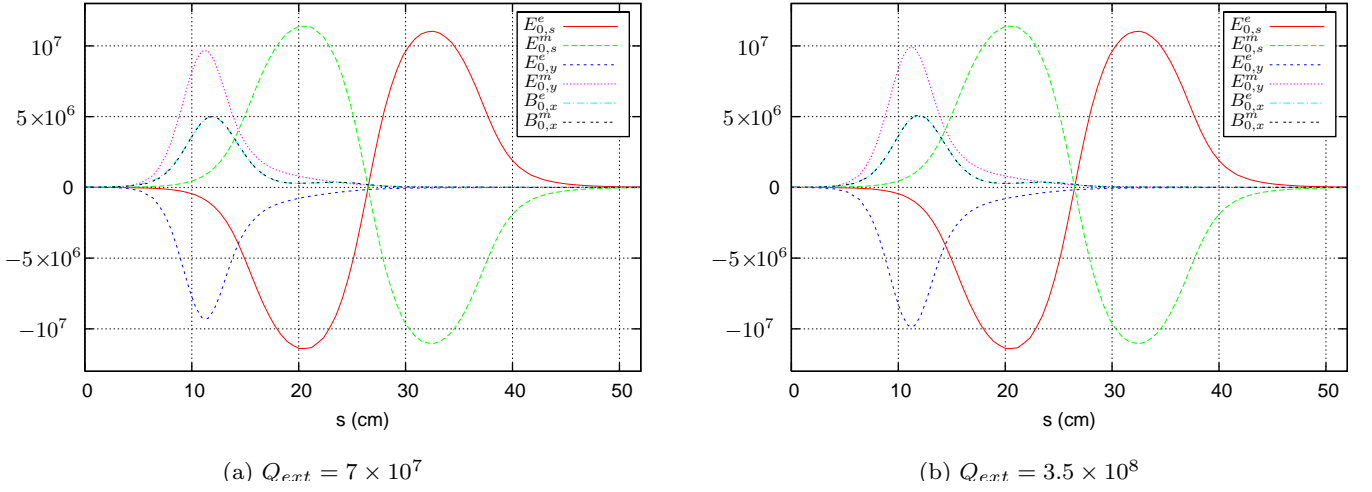}
\caption{Field profiles of MWS standing waves for $Q_{ext}$ values of $7\times 10^7$ and $3.5\times 10^8$.
Units are MV/m and T for the electric and magnetic fields respectively.
$E^e_{0,y}$ and $E^m_{0,y}$ are scaled by $10^3$ and $B^e_{0,x}$ and $B^m_{0,x}$ are scaled by $10^9$.  The standing wave approximation is justified with 
$\mathbf{E}_0^m\approx-\mathbf{E}_0^e$ and $\mathbf{B}_0^m\approx\mathbf{B}_0^e$
\label{fig2}}
\end{figure*}

\subsection{Obtaining Realistic $Q_{ext}$ Values}
Simulations in MWS were run varying the depth of the inner conductor in order to obtain $Q_{ext}$ values
in the vicinity of the proposed value $10^8$ \cite{Liepe}.  
In order to obtain the high $Q_{ext}$ values it 
is necessary to raise the inner conductor into the coupler, signified by a negative depth value.  
The depth used to achieve two high $Q_{ext}$ values 
are $-9.6$ mm for $Q_{ext}=7\times 10^7$ and $-16.4$ mm for $Q_{ext}=3.5\times 10^8$.  
The field profiles along the central cavity axis are shown in Fig. \ref{fig2}.  
For these calculations the coupler boundary is positioned such that $s^e=-1$.  From these profiles it is clear that 
$\mathbf{E}_0^m\approx-\mathbf{E}_0^e$ and $\mathbf{B}_0^m\approx\mathbf{B}_0^e$
and that therefore the standing wave approximation is justified for these large $Q_{ext}$ 
values.

\section{Calculation of Coupler Kick}
In this section we present the methods used to calculate a realistic value for the coupler kick.  The calculation involves integration of the 
synthesized field profiles simulated in MWS to get the total, complex change in momentum of one charged particle.  In addition we 
use the standing wave approximation to support analytically our results of emittance growth from simulation of one bunch of electrons through the Cornell ERL.  

\subsection{Single Cavity}
From the synthesized waves along the central cavity axis we can determine the Lorentz force on a particle of charge $q$  
traveling down the center
of the cavity,  at velocity $v$, at each position as a function of time and integrate to obtain the total change in momentum. 
We begin with examining the kick due to solely inward traveling waves.  This calculation is equivalent to a cavity with
beam loading with a negligible reflection coefficient:
\begin{equation}
\Delta \mathbf{P}^+ = q\int_{t_{i}}^{t_{f}}[\mathbf{E}^+_0(s,t)+v\mathbf{e}_s\times\mathbf{B}^+_0(s,t)]dt,
\label{eq15}
\end{equation}
with $s=vt$.  
With length $L$ of the cavity we can change the variable of integration to $s$:
\begin{equation}
\Delta \mathbf{P}^+ = \frac{q}{v}\int_{0}^{L}[\mathbf{E}_0^+(s,s/v)+v\mathbf{e}_{s}\times\mathbf{B}_0^+(s,s/v)]ds.
\label{eq16}
\end{equation}
Equation (\ref{eq10a}) leads to
\begin{align}
\label{eq17}
\Delta\mathbf{P}^+ =&e^{i\phi^+}\frac{q}{v}\int_{0}^{L}\{[\xi\mathbf{E}^{m}_0(s)+i\mathbf{E}^{e}_0(s)] \\
                      &+ v\mathbf{e}_s \times[\mathbf{B}^e_0(s)+i\xi\mathbf{B}^m_0(s)]\}e^{-i\omega \frac{s}{v}}ds.\nonumber
\end{align}
From now on, as in the above equation, we will work with complex expressions for the change in momentum of which
the real part is physical. 
For an electron arriving at $s=0$ at a time $\Delta t$ the kick is obtained
by replacing $s$ with $s+v\Delta t $ in the exponent of Eq.~(\ref{eq17}).  

The coupler kick $\kappa$ is defined as the ratio of the transverse kick and the longitudinal accelerating kick.  
Defining our axes such that the transverse kick resides solely in the $y$ direction we have for the coupler kick
\begin {equation}
\kappa^+=\frac{\Delta P^+_y}{\Delta P^+_s}= \frac{|\Delta P^+_y|}{|\Delta P^+_s|}
e^{i\phi_c}
\label{eq18}
\end{equation}  
where
\begin{equation}
\Delta P_{y}^+=\frac{e}{c}\int_{0}^{L}[E_{0,y}^+(s,s/c)+cB_{0,x}^+(s,s/c)]ds
\label{eq19}
\end{equation}
and
\begin{equation}
\triangle P_{s}^+=\frac{e}{c}\int_{0}^{L}E_{0,s}^+(s,s/c)ds.
\label{eq20}
\end{equation}

\subsection{Effect due to Alternating Position of Coupler}
In the MWS simulations the coupler is situated in front of the cavity.  However, in configurations (af) and (aa) the
coupler will alternate from being placed in front of and behind the cavity.  It is therefore necessary
to model the change in momentum due to a coupler kick supplied after the particle exits the cavity.  We find that 
the same MWS field profiles from the simulations with the coupler in front of the cavity can be used for this second 
calculation.  The transverse fields with the alternate position of the coupler 
can be modeled by taking the mirror image of the original fields, negating the magnetic field so 
as to ensure the traveling wave in the coax satisfies Maxwell's equations. From Eq.~(\ref{eq17}) the subsequent transverse
and longitudinal kicks can be written as
\begin{align}
\label{eq21}
&\Delta\mathbf{P}_{\bot}'^+ =e^{i\phi^+}\frac{q}{v}\int_{0}^{L}\{[\xi\mathbf{E}^{m}_{0,\bot}(L-s)+i\mathbf{E}^{e}_{0,\bot}(L-s)] \\
                      &- v\mathbf{e}_s \times[\mathbf{B}^{e}_{0,\bot}(L-s)+i\xi\mathbf{B}^{m}_{0,\bot}(L-s)]\}e^{-i\omega \frac{s}{v}}ds\nonumber
\end{align}
and
\begin{equation}
\label{eq21a}
\Delta P_{s}'^+=-e^{i\phi^+}\frac{q}{v}\!\int_{0}^{L}\!\![\xi E^{m}_{0,s}(L-s)+iE^{e}_{0,s}(L-s)]e^{-i\omega \frac{s}{v}}ds.
\end{equation}
A change of variables from $s$ to $L-s$ leads to
\begin{align}
\label{eq22}
\Delta\mathbf{P}_{\bot}'^+ =&e^{i\phi^+}\frac{q}{v}\int_{0}^{L}\{[\xi\mathbf{E}^{m}_{0,\bot}(s)+i\mathbf{E}^{e}_{0,\bot}(s)] \\
                      &- v\mathbf{e}_s \times[\mathbf{B}^e_{0,\bot}(s)+i\xi\mathbf{B}^m_{0,\bot}(s)]\}e^{i\omega \frac{s-L}{v}}ds\nonumber
\end{align}
and
\begin{equation}
\label{eq22a}
\Delta P_{s}'^+=-e^{i\phi^+}\frac{q}{v}\!\int_{0}^{L}\!\![\xi E^{m}_{0,s}(s)+iE^{e}_{0,s}(s)]e^{i\omega \frac{s-L}{v}}ds.
\end{equation}
We now compare the coupler kicks due to the two different positions of the coupler, starting with the expressions for
change in momentum of Eqs.~(\ref{eq17}), (\ref{eq22}) and (\ref{eq22a}).  We will restrict the analysis to highly
relativistic particles.  Making the substitutions of Eq.~(\ref{eq14}), the expression
for the change in momentum of Eq.~(\ref{eq17})   
simplifies to
\begin{align}
\label{eq23}
\Delta\mathbf{P}^+ &\approx (\xi+is^e)e^{i\phi^+}\frac{q}{c}\int_{0}^{L}\{\mathbf{E}^{m}_0(s)\hspace{2cm} \\
              &\hspace{3cm}+ i c\mathbf{e}_s \times\mathbf{B}^m_0(s)\}e^{-i\omega \frac{s}{c}}ds \nonumber\\
	      &\approx A\mathbf{F} \nonumber
\end{align}
with $A$ defined in Eq.~(\ref{eq13a}).
Similarly for Eqs.~(\ref{eq22}) and (\ref{eq22a}):
\begin{align}
\label{eq24}
\Delta\mathbf{P}_{\bot}'^+&\approx (\xi + is^e)e^{i\phi^+}\frac{q}{c}\int_{0}^{L}\{\mathbf{E}^{m}_{0,\bot}(s)\hspace{2cm} \\
                      &\hspace{3cm}-i c\mathbf{e}_s \times\mathbf{B}^m_{0,\bot}(s)\}e^{i\omega \frac{s-L}{c}}ds \nonumber\\
		   &\approx A\mathbf{F}_{\bot}^*e^{-i\omega\frac{L}{c}}\nonumber
\end{align}
and
\begin{align}
\label{eq24a}
\Delta P_s'^+&\approx -(\xi + is^e)e^{i\phi^+}\frac{q}{c}\int_{0}^{L}E^{m}_{0,s}(s)e^{i\omega \frac{s-L}{c}}ds \\
		 &\approx -AF_{s}^*e^{-i\omega\frac{L}{c}}.\nonumber
\end{align}
Evaluating the coupler kicks through substitution into Eq.~(\ref{eq18}) leads to cancellation of the constant terms $\xi + is^e$
along with the exponential
term $e^{-i\omega \frac{L}{c}}$ in the expression for $\Delta P'^+ $.  We thus obtain for the two coupler kicks
\begin{equation}
\kappa^+\approx \frac{\int_{0}^{L}\{\mathbf{E}^{m}_{0,y}(s)+ i c\mathbf{e}_s \times\mathbf{B}^m_{0,x}(s)\}e^{-i\omega \frac{s}{c}}ds}
		  {\int_{0}^{L}\mathbf{E}^{m}_{0,s}(s)e^{-i\omega \frac{s}{c}}ds}=\frac{F_y}{F_s}
\label{eq25}
\end{equation}
and
\begin{equation}
\kappa'^+\approx \frac{\int_{0}^{L}\{\mathbf{E}^{m}_{0,y}(s)- i c\mathbf{e}_s \times\mathbf{B}^m_{0,x}(s)\}e^{i\omega \frac{s}{c}}ds}
		  {-\int_{0}^{L}\mathbf{E}^{m}_{0,s}(s)e^{i\omega \frac{s}{c}}ds}=-\frac{F^*_y}{F^*_s}.
\label{eq26}
\end{equation}
The result of this comparison is the observation that the coupler kick due to the coupler situated at the end of the cavity 
is the negative complex conjugate of the coupler kick due to a coupler located at the beginning of the cavity:
\begin{equation}
\kappa'^+\approx -(\kappa^+)^*
\label{eq27}
\end{equation}

We can now calculate the approximate effect on emittance and orbit distortion from two consecutive cavities with the couplers placed
before the first cavity and after the second cavity, i.e. configurations (ta) and (aa).  For configuration (ta)
where both coupler kicks are in the same direction this can be done by adding 
to Eq.~(\ref{eq3}) a second similar equation with the coupler phase $\phi_c$ changed to $-\phi_c + \pi$, from 
Eq.~(\ref{eq27}):
\begin{align}
\label{eq28}
\Delta y'&\approx\frac{\Delta E_0}{E}|\kappa|\{\cos(\phi_c+\psi)-\omega\sin(\phi_c+\psi)\Delta t\\
         &\hspace{10 mm}+\cos(-\phi_c+\psi+\pi)-\omega\sin(-\phi_c+\psi+\pi)\Delta t\} \nonumber \\
          &\approx-2\frac{\Delta E_0}{E}|\kappa|\{\sin(\phi_c)\sin(\psi)+\omega\sin(\phi_c)\cos(\psi)\Delta t\}.\nonumber
\end{align}
Similarly we can approximate the effect on emittance from the (aa) configuration by instead subtracting 
the second kick from Eq.~(\ref{eq3}):
\begin{align}
\label{eq28a}
\Delta y'&\approx\frac{\Delta E_0}{E}|\kappa|\{\cos(\phi_c+\psi)-\omega\sin(\phi_c+\psi)\Delta t\\
         &\hspace{10 mm}-\cos(-\phi_c+\psi+\pi)+\omega\sin(-\phi_c+\psi+\pi)\Delta t\} \nonumber \\
          &\approx 2\frac{\Delta E_0}{E}|\kappa|\{\cos(\phi_c)\cos(\psi)-\omega\cos(\phi_c)\sin(\psi)\Delta t\}.\nonumber
\end{align}
On crest operation, or $\psi = 0$, leads to a cancellation of the $\Delta t$ term in Eq.~(\ref{eq28a}) and thus to no 
emittance growth with the (aa) configuration, while for the (ta) configuration on crest operation with $\psi = 0$ 
leads to no orbit distortion $\Delta y_0'$.  Other effects that we can deduce from the above two equations are that 
with 
off-crest operation, $\psi = \pi$ as in a bunch compressor or a hadron storage ring,
there is zero emittance growth with the (ta) configuration and zero orbit distortion with
the (aa) configuration.

Coupler kicks were calculated using Eqs.~(\ref{eq17}) and (\ref{eq22}), without assuming standing wave approximation, by
numerical integration of the field profiles of MWS in MathCad.  The phase of the coupler kicks and the respective 
magnitudes are 
shown in Table \ref{tab1}, for both proposed $Q_{ext}$ values and for the two positions of the coupler, 
before and after the cavity.  
From the results we see that Eq.~(\ref{eq27}) holds: the coupler strengths are equivalent for
the different positions of the coupler and the coupler phases are are related via $\phi_c'=-\phi_c + \pi$.  
The position of the coupler boundary in the MWS simulations were chosen such that 
$\xi=1$ to achieve equivalent accuracies of the electric and magnetic boundary field profiles.  
We have observed that choosing a coupler boundary position with
a very large $\xi$ leads to low accuracy in the electric boundary fields, and choosing a 
position with a small $\xi$ leads to low accuracy in the magnetic boundary fields.

\begin{table}
\caption{Coupler kick parameters.
\label{tab1}}
\begin{tabular}{|l||l|l||l|l|}
\hline
 &\multicolumn{2}{l|}{$Q_{ext}=7\times 10^{7}$}&\multicolumn{2}{l|}{$Q_{ext}=3.5u\times 10^{8}$}\\ 
\cline{2-5}
 &Before Cav&After Cav&Before Cav&After Cav\\
\hline\hline
$|\kappa|(10^{-4})$&.9651&.9891&1.039&1.027\\
$\phi_c$ (rad)& 2.838& 0.349& 2.819&0.326\\
\hline
\end{tabular}
\end{table}

\begin{figure*}[tp]
\includegraphics[width=\textwidth, clip]{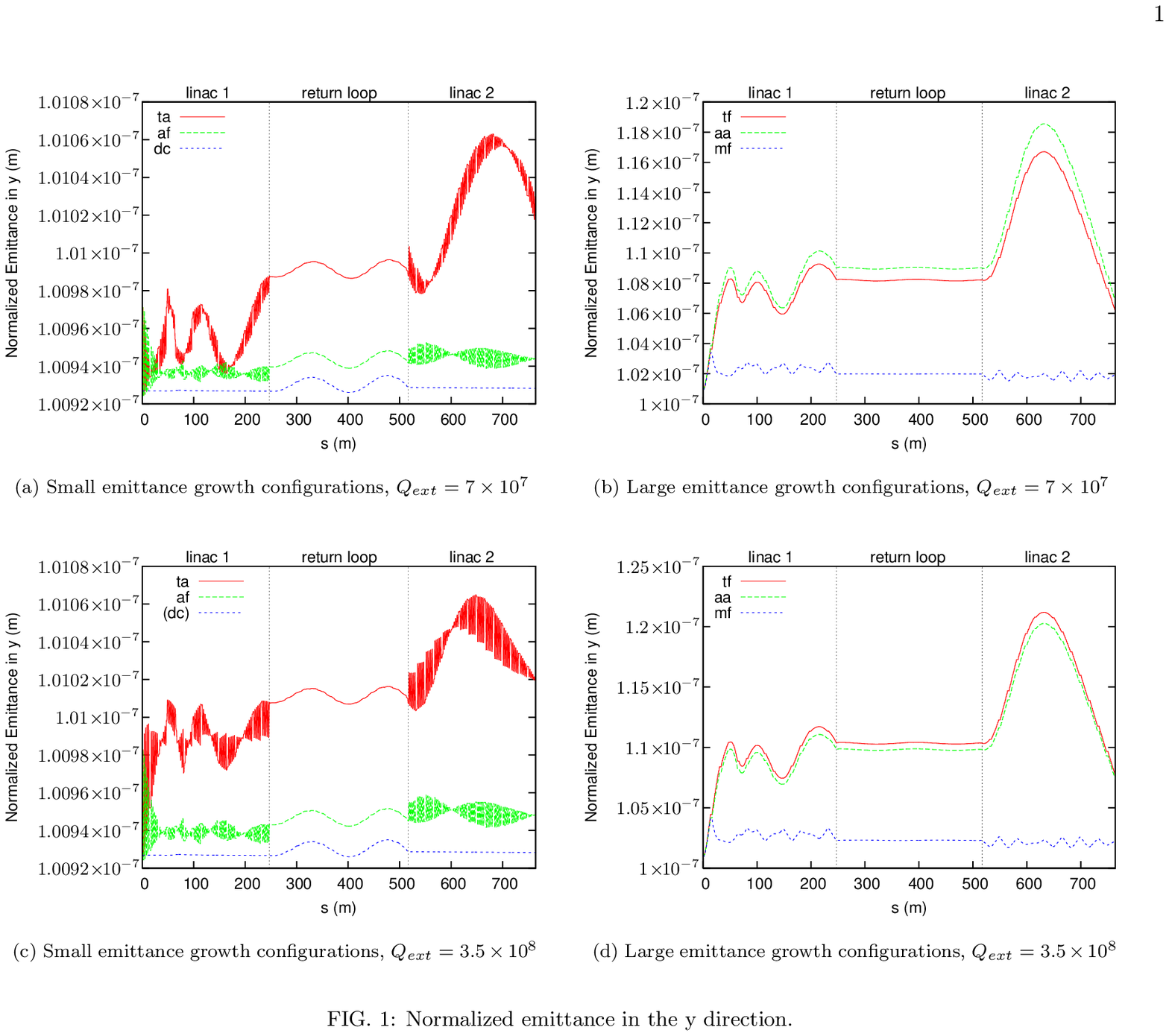}
\caption{Normalized emittance in the y direction.  
\label{fig5}}
\end{figure*}

Shown in Fig.~\ref{fig5} are results of normalized emittance from simulations in BMAD through the ERL 
lattice with the calculated coupler kick values
for both proposed $Q_{ext}$ values and for all six coupler configurations.  
The initial normalized emittance is $1\times10^{-7}$m.  The Cornell ERL is split into 
two accelerating sections, labeled as linac 1 and 2, connected by a return loop \cite{hoffstaetter07_1}.
To compensate for overall transverse kicks, the necessary corrector coil strengths are 
computed and included in the lattice.  

As one might expect from 
our previous conclusion, the increase in normalized emittance is small for the (aa) and (af) configurations
while large for the (ta) configuration which has a nearly identical effect as the (tf) configuration.  
Hence, these $Q_{ext}$ values of $7\times 10^7$ and $3.5\times 10^8$ are large enough to sufficiently
satisfy the standing wave approximation.  We have found 
that $Q_{ext}$ values in the vicinity of $10^5$, such as for the ERL injector cavities, do not satisfy the 
standing wave approximation well enough and the emittance growth is not sufficiently small for the (aa) configuration.
In our experience, the standing wave approximation holds sufficiently well for $Q_{ext}$ values greater than 
$10^7$.

From these results we come
to the conclusion that for operating at or near on crest, configuration (aa) is preferable if conservation of
emittance is of primary concern.  Configuration (ta) is preferable for operating completely off-crest as is 
apparent after substitution of $\psi = \pi$ into Eq.~(\ref{eq28}).  However,
for certain applications minimizing the orbit distortion and hence the overall transverse kick is of importance.
From Eqs.~(\ref{eq28}) and (\ref{eq28a}) we see that the (ta)
configuration results in less of a
transverse orbit distortion than does the (aa) configuration with on crest operation and may be a preferable 
configuration than the (aa)
configuration in certain applications. 


\begin{figure*}[tp]
\includegraphics[width=\textwidth, clip]{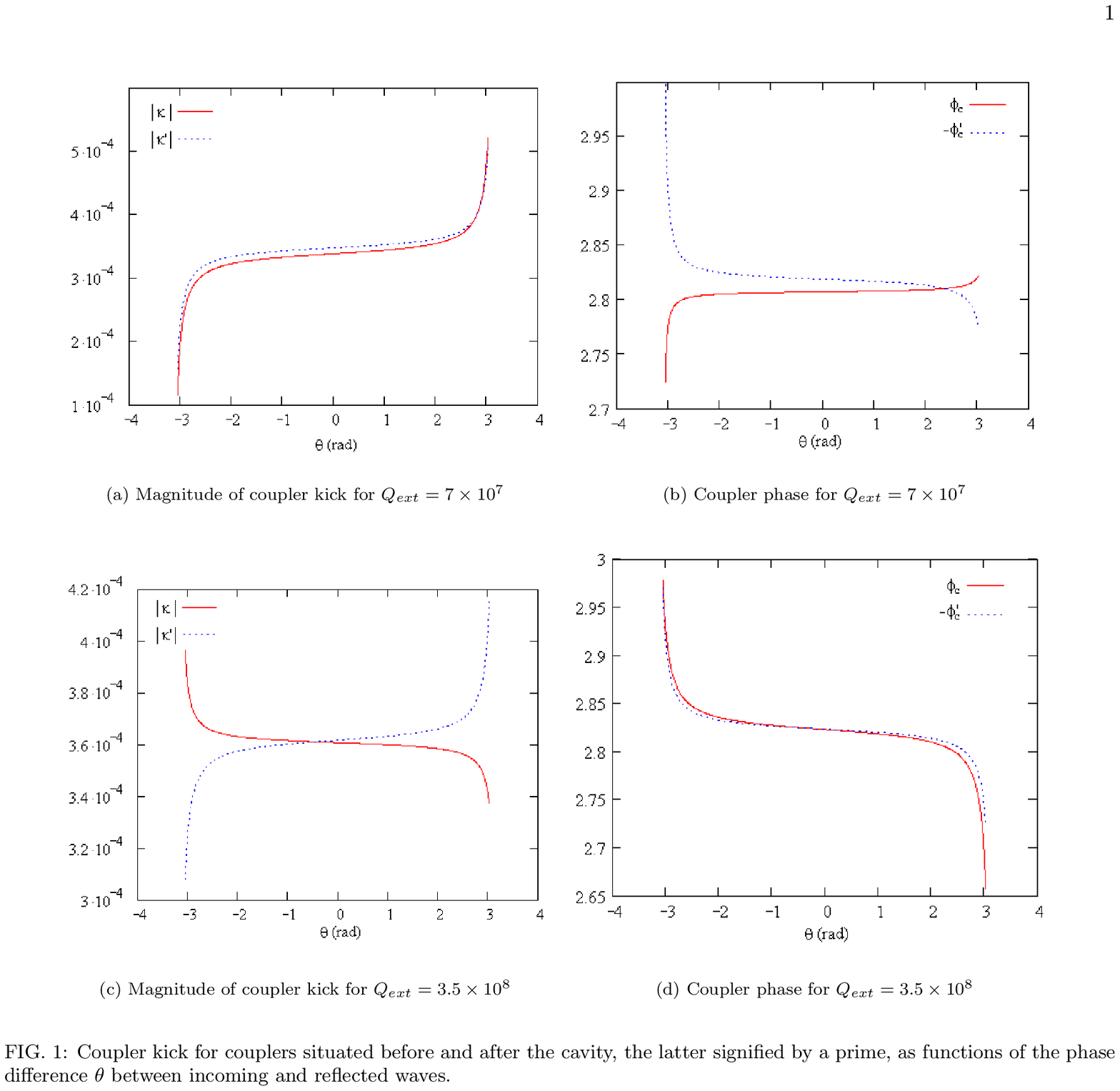}
\caption{Coupler kick for couplers situated before and after the cavity, the latter signified by a prime, as functions
of the phase difference $\theta$ between incoming and reflected waves.  
\label{fig6}}
\end{figure*}

\subsection{Reflected Waves in the Cavity}
In many applications cavities are operated with large reflection of the incoming RF wave.  For example, in an ERL, for which there are an equal number of 
accelerating bunches as there are  decelerating bunches, beam loading can 
be neglected and the incoming energy is not 
transferred to the beam in steady state operation.  Because the value of $Q_0$ is large compared to that of $Q_{ext}$ inside the superconducting cavities, 
nearly all of the incoming energy will be reflected in RF waves traveling back up the coupler.   Both the incoming and the outgoing waves will excite standing 
waves in the cavity.  These standing waves will differ by a phase factor $\phi$ determined by the cavity 
detuning, with a phase difference of zero for on resonance operation.  The amplitudes 
will be equal for full reflection.  It is necessary to examine the coupler kick due to a superposition of incoming and 
outgoing waves and to determine
whether the result of Eq.~(\ref{eq27}), namely the cancellation of emittance growth due to alternating the coupler from front to back of the cavity, 
 still holds for arbitrary phase differences and different detuning of adjacent 
cavities.

Due to the reflected waves, in addition to ${\bf \Delta P^+}$ there will be kicks:
\begin{align}
\label{eq29}
\Delta \mathbf{P}^- &= \frac{q}{v}\int_{0}^{L}[\mathbf{E}_0^-(s,s/v)+v\mathbf{e}_{s}\times\mathbf{B}_0^-(s,s/v)]ds \\
                    &=e^{i\phi^-}\frac{q}{v}\int_{0}^{L}\{[\xi\mathbf{E}^{m}_0(s)-i\mathbf{E}^{e}_0(s)]\nonumber \\
                      &\hspace{4 mm} - v\mathbf{e}_s \times[\mathbf{B}^e_0(s)-i\xi\mathbf{B}^m_0(s)]\}e^{-i\omega \frac{s}{v}}ds\nonumber
\end{align}
with the coupler situated in front of the cavity
and
\begin{align}
\label{eq30}
& \Delta \mathbf{P'}_{\bot}^- =e^{i\phi^-}\frac{q}{v}\int_{0}^{L}\{[\xi\mathbf{E}^{m}_{0,\bot}(s)-i\mathbf{E}^{e}_{0,\bot}(s)]\\ 
& \hspace{4 mm} + v\mathbf{e}_s \times[\mathbf{B}^e_{0,\bot}(s)-i\xi\mathbf{B}^m_{0,\bot}(s)]\}e^{i\omega \frac{s}{v}}e^{-i\omega \frac{L}{v}}ds\nonumber
\end{align}
and
\begin{equation}
\label{eq30a}
\Delta {P'}_s^-=-e^{i\phi^-}\frac{q}{v}\int_{0}^{L}[\xi E^m_{0,s}(s)-i E^e_{0,s}(s)]e^{i\omega \frac{s}{v}}e^{-i\omega \frac{L}{v}}ds.
\end{equation}
for the coupler situated after the cavity.  Making the substitutions of Eq.~(\ref{eq14}) and setting $v=c$ for 
highly relativistic particles leads to:
\begin{align}
\label{eq31}
\Delta\mathbf{P}^-&\approx (\xi - is^e)e^{i\phi^-}\frac{q}{c}\int_{0}^{L}\{\mathbf{E}^{m}_0(s) \hspace{2cm} \\
                      &\hspace{3cm}+ i c\mathbf{e}_s \times\mathbf{B}^m_0(s)\}e^{-i\omega \frac{s}{c}}ds\nonumber\\
		  &\approx A\mathbf{F}\nonumber
\end{align}
and
\begin{align}
\label{eq24b}
\Delta\mathbf{P}_{\bot}'^-&\approx (\xi - is^e)e^{i\phi^-}\frac{q}{c}\int_{0}^{L}\{\mathbf{E}^{m}_{0,\bot}(s)\hspace{2cm} \\
                      &\hspace{3cm}-i c\mathbf{e}_s \times\mathbf{B}^m_{0,\bot}(s)\}e^{i\omega \frac{s-L}{c}}ds \nonumber\\
		   &\approx A\mathbf{F}_{\bot}^*e^{-i\omega\frac{L}{c}}\nonumber
\end{align}
and
\begin{align}
\label{eq24c}
\Delta P_s'^-&\approx -(\xi - is^e)e^{i\phi^-}\frac{q}{c}\int_{0}^{L}E^{m}_{0,s}(s)e^{i\omega \frac{s-L}{c}}ds \\
		 &\approx -AF_{s}^*e^{-i\omega\frac{L}{c}}.\nonumber
\end{align}

The coupler kick $\kappa(\alpha)$, including these reflected waves, is thus given by
\begin {equation}
\kappa(\alpha)=\frac{\Delta P^+_y+\alpha\Delta P^-_y}{\Delta P^+_s+\alpha\Delta P^-_s},
\label{eq33}
\end{equation} 
where $\Delta \mathbf P^+$ and $\Delta \mathbf P'^+$ are given in Eqs.~(\ref{eq23}), (\ref{eq24}) 
and (\ref{eq24a}) and $\alpha$ is the complex reflection coefficient.  

We can now compare the coupler kicks including the reflected waves from a coupler situated in front of the cavity and a coupler situated after the cavity:  
\begin{equation}
\kappa(\alpha)\approx \frac{AF_y+\alpha A F_y}{A F_s+\alpha A F_s}=\frac{F_y}{F_s}= \kappa^+
\label{eq34}
\end{equation}
and
\begin{equation}
\kappa'(\alpha')\approx \frac{A F^*_y+\alpha' A F^*_y}{-A F^*_s-\alpha' A F^*_s}=\frac{F^*_y}{-F^*_s}=\kappa'^+=-(\kappa^+)^*.
\label{eq35}
\end{equation}
We find that the coupler kick is independent of reflected waves and their phases relative to the incoming waves and thus 
again the complex conjugate relationship should be valid for arbitrary detuning:   
\begin{equation}
\kappa'\approx -\kappa^*
\label{eq36}
\end{equation}
for any values of $\alpha$ and $\alpha'$.  Therefore, the orbit distortion from two successive cavities for which 
the couplers are on different sides of their respective
cavities but mounted on the same side of the beam pipe still cancel even with reflection.  

\unitlength=1cm
\begin{picture}(0,0)
\put(14.5,12.52){$\pi$}
\end{picture}

\unitlength=1cm
\begin{picture}(0,0)
\put(14.55,20.67){$\pi$}
\end{picture}

Figure \ref{fig6} plots, for both proposed $Q_{ext}$
values, the phase and amplitude of the coupler kicks as a function of
the phase difference between incoming and reflected waves for two
adjacent cavities with full reflection, $|\alpha|=1$.  As before, the position of the boundary is chosen
with $\xi=1$ and $s^e=-1$.  The phase difference $\theta$ is varied from $-\pi$ to $\pi$ with a phase difference of zero for no detuning.  For small detuning,
with a phase difference around 0,
the negative complex conjugacy approximation is satisfied very well.

\section{Alternative Methods for Reducing Coupler-Kick Effect}
\subsection{Minimizing Coupler Phase}
As illustrated previously, the alternating phase of the coupler kick due to the alternating placement of the coupler leads to low emittance growth
and/or lower orbit distortion.
An alternative method for minimizing emittance growth which does not depend on the alternating placement of the coupler 
entails manipulating the coupler kick such that its phase is $0$ or $\pi$.  As the change in emittance of Eq.~(\ref{eq6}) varies with
$S^2$ and thus with $\sin^2(\phi_c+\psi)$, operation at $\psi=0$ leads to low  emittance growth 
for $\phi_c=0$ or $\pi$.  This method reduces the effects from each individual coupler and is effective no matter the configuration of couplers 
along the lattice.      

The coupler kick phase is dependent on the distance the coupler is situated from the entrance of the cavity.
In the previous simulations the coupler was positioned 4.5 cm from the entrance of the cavity.  
We find that moving the coupler out to a distance of 5.3 cm leads to a coupler phase of $\pi$ for $Q_{ext}=7\times10^7$
and moving out to a distance of 5.5 cm leads to a phase of $\pi$ for $Q_{ext}=3.5\times 10^8$.  The coupler kick parameters are listed in 
Table \ref{tab2}. 

\begin{table}
\caption{Coupler kick parameters with optimized coupler phase.
\label{tab2}}
\begin{tabular}{|l||l|l||l|l|}
\hline
 &\multicolumn{2}{l|}{$Q_{ext}=7\times 10^{7}$}&\multicolumn{2}{l|}{$Q_{ext}=3.5\times 10^{8}$}\\
\cline{2-5}
 &Before Cav&After Cav&Before Cav&After Cav\\
\hline\hline
$|\kappa|$($10^{-4})$&.6037&.6066&.5943&.6043\\
$\phi_c$(rad)& 3.126&0.129&3.129&0.042\\
\hline
\end{tabular}
\end{table}

\begin{figure}[tp]
\includegraphics[width=\columnwidth, clip]{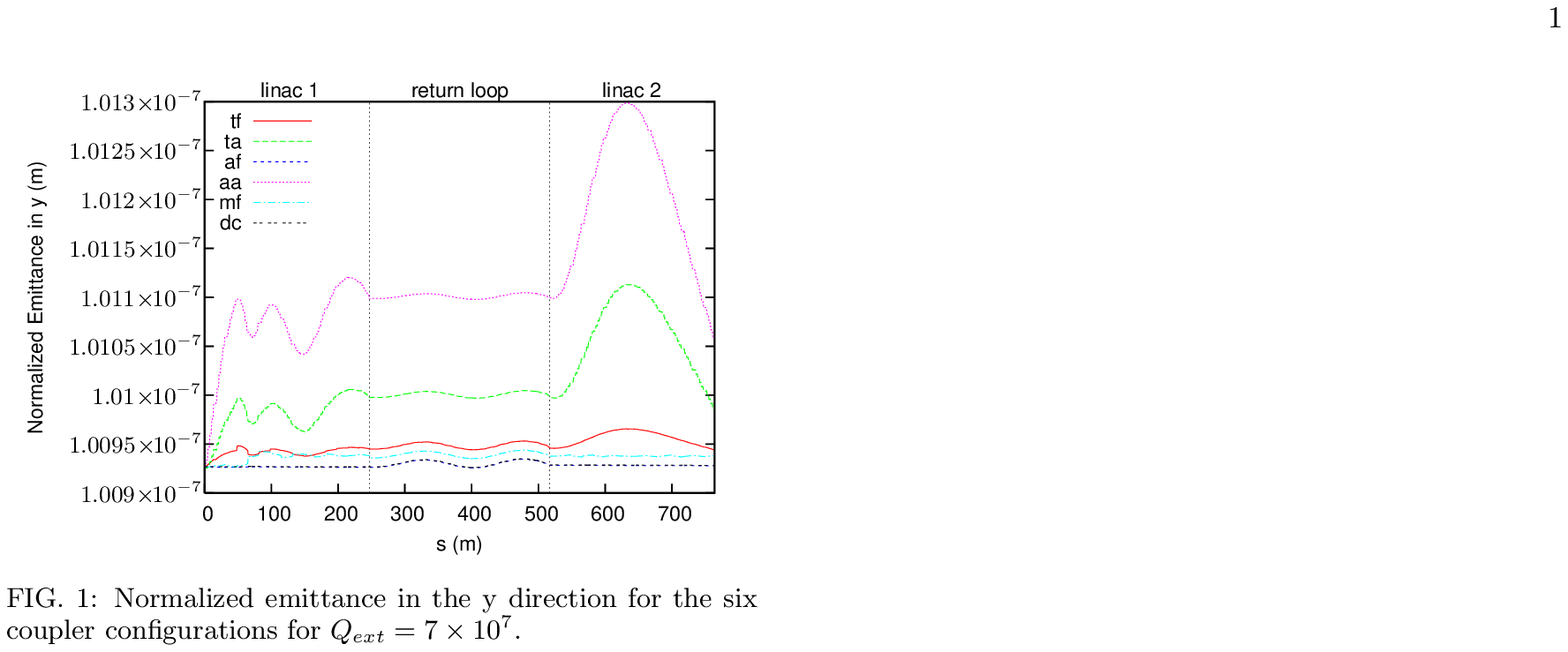}
\caption{Normalized emittance in the y direction for the six coupler configurations for $Q_{ext}=7\times 10^{7}$.
\label{fig8}}
\end{figure}

\begin{figure}[tp]
\includegraphics[width=\columnwidth, clip]{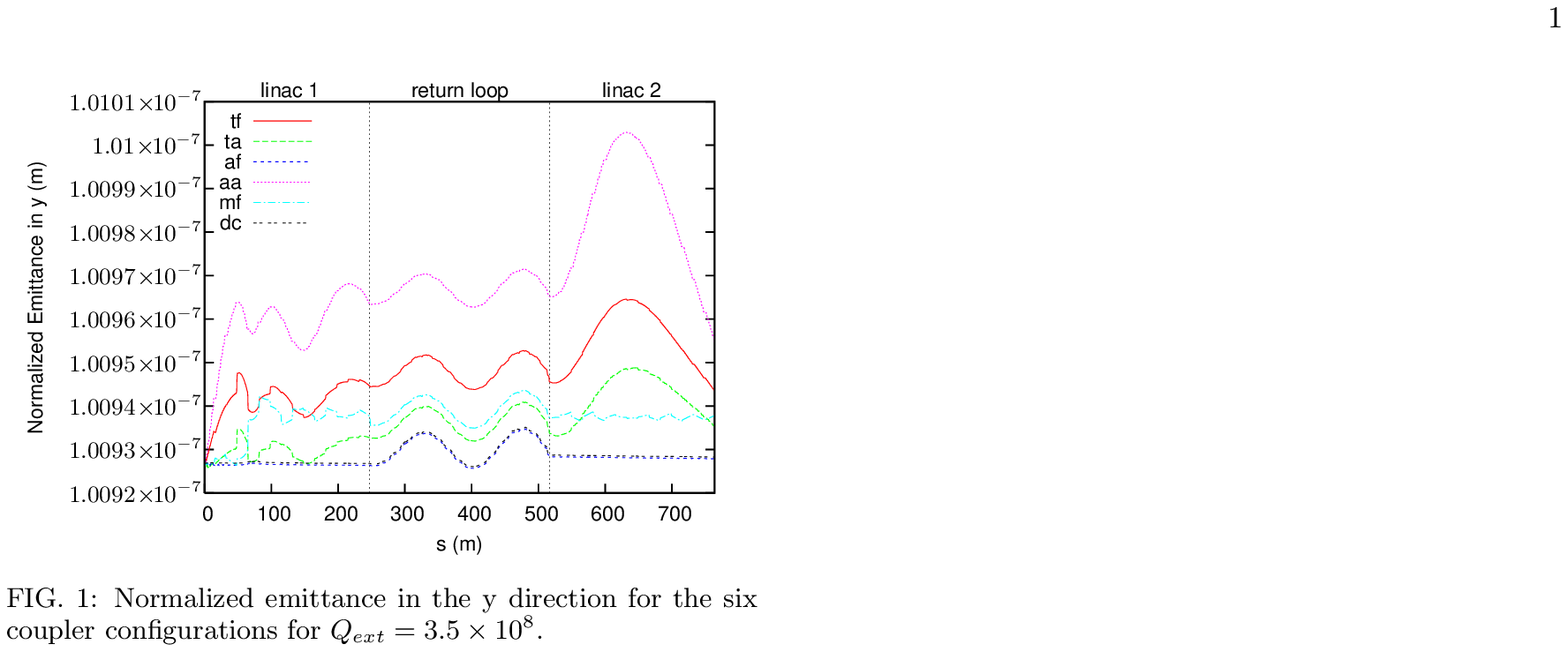}
\caption{Normalized emittance in the y direction for the six coupler configurations for $Q_{ext}=3.5\times 10^{8}$.
\label{fig9}}
\end{figure}

Figures \ref{fig8} and \ref{fig9} show the results of normalized emittance through the ERL lattice 
for all six coupler arrangements with the coupler parameters of Table \ref{tab2}.  
The emittance growth is decreased
substantially for all cases illustrating the dependence of the emittance growth on the phase of the coupler kick.    

\subsection{Symmetrizing Stub}
\begin{figure}[tbhp]
\includegraphics[width=\columnwidth, clip]{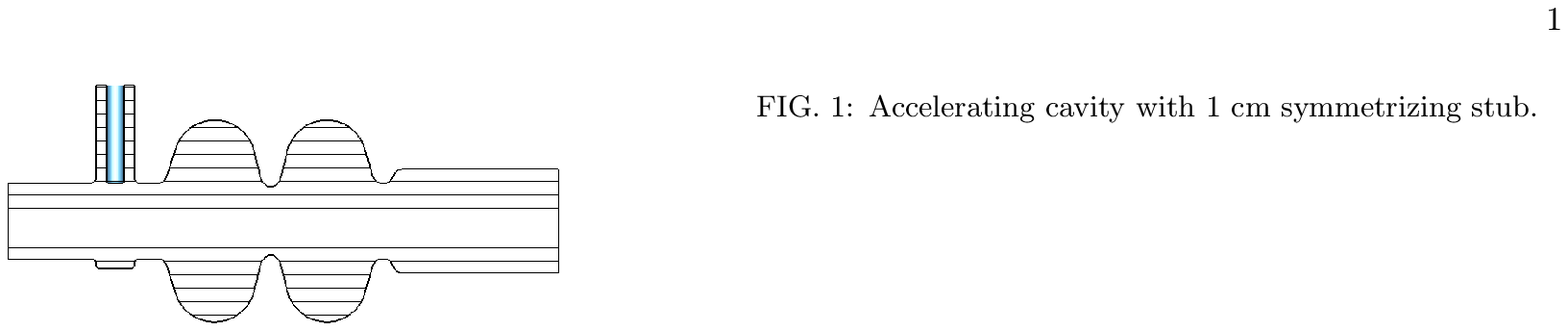}
\caption{Accelerating cavity with 1 cm symmetrizing stub. 
\label{fig10}}
\end{figure}

The above methods for reducing emittance growth, namely the alternating position of the coupler as 
in configuration (aa) and the phase minimization technique, 
all depend on operation on crest, $\psi=0$.  For certain applications it is preferable to operate 
slightly off crest.  For such applications 
an alternative method for reducing emittance growth is adding a stub across 
from the coupler as illustrated in Fig. \ref{fig10}.  The stub is used to minimize the asymmetry in the beam pipe
causing the transverse fields in the coupler region.  The method reduces amplitudes of the off 
axis fields and thus reduces the 
magnitude of the coupler kick depending on
the depth of the stub, a larger stub leading to lower off axis field amplitudes.

Simulations were run with configuration (aa) $9^{\circ}$ off crest with the coupler placed 4.5 cm from the entrance 
of the beam pipe, i.e.~phase not minimized, to investigate the extent of the dependence of the emittance growth cancellation 
on $\psi$.  A second simulation was run with the same configuration, $\psi=9^{\circ}$, but with a stub 
of only 1 cm depth added to the cavity.  The 1 cm depth is not the 
result of an optimization but is chosen small enough
so as to illustrate the effectiveness of the symmetrizing stub.  Larger stub depths did not result in 
less emittance growth.  

\begin{figure}[bhp]
\includegraphics[width=\columnwidth, clip]{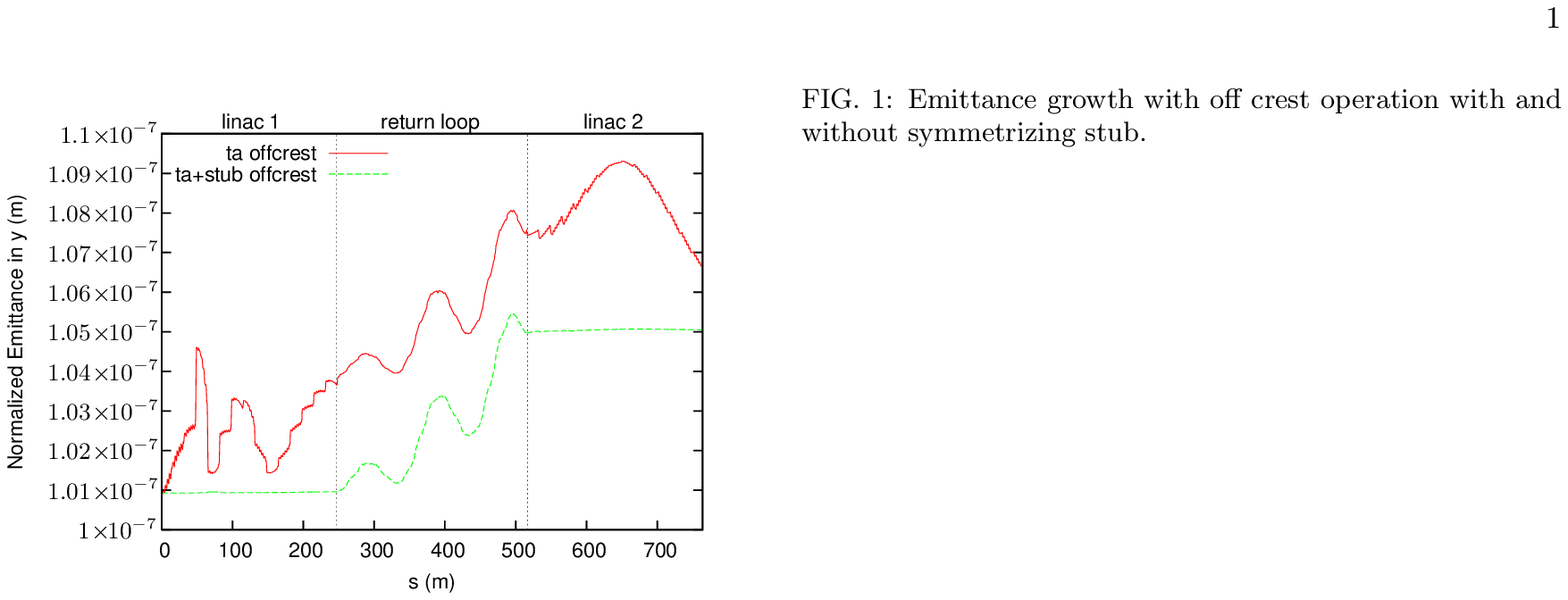}
\caption{Emittance growth with off crest operation with and without symmetrizing stub.  
\label{fig11}}
\end{figure}

As illustrated in Fig. \ref{fig11}, the emittance growth with no stub is significantly larger 
than the previous, on crest simulations, Fig.~\ref{fig5},
illustrating the dependence on $\psi$.  The addition of the only 1cm long stub eliminates emittance growth 
through the two linacs very effectively.  The emittance increase in the
return loop between linacs is independent of the coupler kicks.   
 
\section{Conclusion}
We have investigated three methods of minimizing the emittance growth due to coupler kicks in linacs: 
(a) alternating the position and direction of the coupler each cavity, (aa) configuration, 
(b) Choosing the distance between coupler and cavity to minimize the coupler 
kick for on crest acceleration, (c) symmetrizing 
the coupler region by adding a stub opposite the coupler.  All three methods are shown to work 
very well.  For (a) we find that it is necessary to implement the more technically challenging configuration
of alternating the side of the beam pipe the coupler is mounted each cavity.  However, we find that for
techniques (b) and (c) the one-sided coupler configurations (tf) and (ta) lead to sufficiently low emittance
growth.  
For technique (c) it is interesting to note that a very small symmetrizing stub of only 
1 cm can suppress emittance growth very well., independent of the acceleration phase.
In addition, method (c) produces very small orbit distortions, similar to configuration (ta) and (af) which do not have small emittance growth.

\section*{Acknowledgments}
The authors wish to thank Joseph Choi for their initial collaboration, Valery Shemelin 
for his previous studies on the subject and helpful guidance, 
David Sagan for sharing his vast wisdom of BMAD and Richard Helms for his critiques and 
suggestions on presentation of results. We thank Martin Dohlus for pointing out a sign error in \cite{hoffstaetter07_2}, which has been corrected in this paper.  This work has been supported by NSF Cooperative Agreement No.~PHY-0202078.


\begin{thebibliography}{10}

\bibitem{Zhang} M. Zhang and Ch. Tang, \textit{Beam Dynamics of the Tesla Power Coupler}, PAC99, New York/NY (1999)

\bibitem{Dohlus} M. Dohlus, \textit{The Influence of the Main-Coupler Field on the Transverse Emittance of a Superconducting RF Gun}, EPAC04, Lucerne/CH (2004)

\bibitem{Belomestnykh02} S. Belomestnykh, M. Liepe, H. Padamsee,
V. Shemelin and V. Veshcherevich, High Average Power Fundamental Input
Couplers for the Cornell University ERL: Requirements, Design
Challenges and First Ideas, Report ERL 02-8 (2002)

\bibitem{shemelin} V. Shemelin, S. Belomestnykh, H. Padamsee,
\textit{Low-Kick Twin-Coaxial and Waveguide-Coaxial Couplers for ERL},
Cornell University Report SRF 021028-08 (2002)

\bibitem{Greenwald} Z. Greenwald and D. L. Rubin, \textit{Emittance Growth Study Using 3DE Code for the ERL Injector Cavities with Various Coupler Configurations},
Cornell University Report ERL 03-09 (2003)

\bibitem{shemelin3} V. Shemelin, S. Belomestnykh, R.L. Geng, M. Liepe. H. Padamsee, Dipole-Mode-Free and Kick-Free 2-Cell Cavity for the SC ERL injector, in Proceedings PAC03 Portland/OR (2003)

\bibitem{MWS}CST Microwave Studio, User Guide, CST GmbH, Budinger Str. D-64289 Darmstadt, Germany (2007)

\bibitem{sagan} D. Sagan, \textit{BMAD Manual}

\bibitem{Dohlus2} M. Dohlus and S. G. Wipf, \textit{Numerical Investigation of Waveguide Input Couplers for the Tesla Superstructure},
EPAC00, Vienna/At (2000)

\bibitem{Hartung} W. Hartung and E. Haebel, \textit{Search of trapped modes in the single-cell cavity prototype for CESR-B}, PAC93, Washington DC (1993)

\bibitem {Balleyquier} P. Balleyguier, \textit{External Q Studies for APT SC-Cavity Couplers}, LINAC'98 Chicago, IL (1998)

\bibitem {Balleyquier2} P. Balleyguier, Particle Accelerators, v. 57, pp. 113-127 (1997)

\bibitem{shemelin2}V. Shemelin, S. Belomestnykh, \textit{Calculation of the B-cell cavity external Q with MAFIA  and Microwave Studios}, Cornell University LNS report SRF 020620-03 (2002)

\bibitem {Kroll} N. Kroll, \textit{Computer Determination of the External Q and Resonant Frequency of Waveguide Loaded Cavities} Particle Accelerators, v. 34, p. 234 (1990)

\bibitem{Liepe} M. Liepe, S. Belomestnykh, J. Dobbins, R. Kaplan, C. Strohman, B. Stuhl, C. Hovater, T. Plawski, \textit{Pushing the Limits: RF Field Control at High Loaded Q}, PAC05, Knoxville/TN (2005)

\bibitem{hoffstaetter07_1} G.H. Hoffstaetter, I.V. Bazarov, D.H. Bilderback, J. Codner, B. Dunham, D. Dale, K. Finkelstein, M. Forster, S. Greenwald, S.M. Gruner, Y. Li, M. Liepe, C. Mayes, D. Sagan, C.K. Sinclair, C. Song, A. Temnykh, M. Tigner, Y. Xie, Proceedings PAC07, Albuquerque/NM (2007)

\bibitem{hoffstaetter07_2} Controlling coupler-kick emittance growth in the Cornell ERL main linac
  G.H. Hoffstaetter, B. Buckley, Proceedings PAC07, Albuquerque/NM (2007)

\end{thebibliography}
\end{document}